\documentclass[conference]{IEEEtran}

\usepackage{booktabs}
\usepackage{tabularx}
\usepackage{array}
\usepackage{multirow}
\usepackage{longtable}
\usepackage{xcolor}
\usepackage{enumitem}
\usepackage{graphicx}
\usepackage{amsmath}
\usepackage{tcolorbox}
\usepackage[hidelinks]{hyperref}

\title{The Transparency Trap: How AI Disclaimers Create Overconfidence
in High-Stakes Decisions\\[0.35em]
\large An Exploratory Study Across Finance, Medicine, and AI-Generated Information}

\author{
\IEEEauthorblockN{Neil Todkar}
\IEEEauthorblockA{
Amador Valley, Class of 2028\\
Pleasanton, CA, USA\\
neiltodkar@gmail.com}
}

\begin{document}

\maketitle

% ─ Abstract ──────────────────────────────────────────────────────────────────
\begin{abstract}
Current AI disclaimers often fail to function as intended due to warning habituation and a transparency paradox. As AI-generated information becomes pervasive in everyday decision-making, effective risk communication is increasingly critical for responsible design. This exploratory study examines how disclaimer placement and persuasive cues shape trust, perceived accuracy, and disclaimer engagement across three high-stakes domains: finance, medicine, and AI-generated content. Using a mixed within-between experimental design with 378 stimulus-level responses from 52 participants, we find that advisory content was generally trusted across conditions, even when disclaimers were present. A significant domain effect showed that medical content received the highest trust ratings. In the AI domain, the findings reveal a transparency paradox: some participants interpreted disclaimers not as warnings, but as signs of system self-awareness and honesty, paradoxically increasing perceived trustworthiness. Evidence of banner blindness further suggests that standardized AI disclaimers are insufficient to prevent over-reliance. Finance and medicine provide useful comparison domains by showing how users interpret warnings differently depending on context and perceived risk. These findings have vital implications for responsible AI design, algorithmic fairness, and consumer protection when users act on potentially misleading information in high-stakes settings.
\end{abstract}


\section{Introduction}
Artificial intelligence is no longer a tool used only by technical specialists. Millions of people now rely on AI-powered systems to answer health questions, evaluate financial decisions, plan travel, and navigate complex personal situations. In many of these contexts, AI responses are accompanied by disclaimers that warn users that the information may be incomplete, inaccurate, or not a substitute for professional advice. These disclaimers are intended to manage user expectations and reduce over-reliance, yet their actual effectiveness remains uncertain. Users repeatedly exposed to similar warnings may stop noticing them, while others may interpret such statements as signs of transparency or self-awareness, making the system appear more trustworthy rather than less. In high-stakes settings, where inaccurate information can have serious consequences, static disclaimers may therefore provide inadequate protection~\cite{henestrosa2025aidisclaimers, benda2021appropriatetrust}.

This exploratory study examines how disclaimers function in practice, focusing on AI-generated information while using finance and medicine as comparison domains. These domains share a common structure: an information source, an advisory warning, and a user who must decide how much to trust and act on the information provided. Prior research shows that warning effectiveness depends on placement, salience, and framing~\cite{wogalter2002warningdesign, laughery2014warningmodel}. However, AI systems introduce a distinct challenge because the same standardized disclaimers often appear repeatedly across many types of responses. Over time, this repetition may produce habituation or banner blindness. At the same time, language such as ``this may be inaccurate'' may be interpreted as a sign of honesty rather than caution. Understanding when and why AI disclaimers fail is therefore important for responsible AI design, user protection, and accountability in cases where users act on incorrect or misleading information.

\section{Related Work}

Prior research on warning design shows that disclaimers are effective only when users notice, understand, and treat them as relevant to the decision being made. Warning salience, placement, formatting, and signal words all influence whether advisory messages attract attention~\cite{wogalter2002warningdesign}. However, exposure alone does not guarantee compliance. Users may see a disclaimer but discount it when the surrounding message appears useful, authoritative, or persuasive. Models of warning compliance therefore emphasize a sequence of noticing, comprehension, and behavioral response~\cite{laughery2014warningmodel}. This framework motivates the present study's focus on disclaimer placement and visual prominence.

Finance and medicine provide important comparison domains because both rely on disclaimers to communicate uncertainty and reduce inappropriate reliance. In medical contexts, warnings may clarify that information is not professional advice, yet users may still trust the content if it appears authoritative or personally relevant. Risk communication research shows that warning effectiveness depends not only on the wording of the disclaimer but also on users' prior expectations, perceived risk, and domain trust~\cite{fischhoff2011riskcommunication}. Similarly, financial disclosure research suggests that users often overlook standardized warnings and instead rely on heuristics such as perceived authority, confidence, or social proof~\cite{sec2013riskdisclosure, green2012mandatorydisclaimers}.

AI-generated information introduces an additional challenge because disclaimers are often repeated across many outputs regardless of topic, confidence level, or risk. Repeated exposure to similar warnings can produce warning fatigue, in which users habituate to advisory messages and begin treating them as routine background text~\cite{cranor2008humanwarnings}. In AI settings, disclaimers may also produce a paradoxical effect: statements such as ``this may be inaccurate or incomplete'' may be interpreted as signs of transparency, honesty, or system self-awareness rather than as cautionary warnings. Prior research on AI trust suggests that users may continue to rely on algorithmic outputs even when uncertainty is explicitly communicated~\cite{benda2021appropriatetrust, scharowski2023explanations, yu2025aitrustethics}. Recent work on AI disclaimers further shows that such warnings can shape trust in ways not always intended by system designers~\cite{henestrosa2025aidisclaimers}.

Together, this literature suggests that disclaimer effectiveness depends on presentation, domain context, repeated exposure, and competing persuasive cues. The present study builds on this work by experimentally comparing disclaimer placement and persuasion conditions across finance, medicine, and AI-generated information.

\section{Research Questions and Hypotheses}

This study examines whether disclaimer placement, domain context, and persuasive cues affect user trust, perceived correctness, and reported disclaimer influence across finance, medicine/wellness, and AI-generated information. The study is guided by four research questions.

\textbf{RQ1: Disclaimer Placement.} Does the placement or visual prominence of a disclaimer affect how much users report being influenced by it?

\textbf{RQ2: Domain Differences.} Does trust in advisory information vary across finance, medicine/wellness, and AI-generated information?

\textbf{RQ3: Persuasion Cues.} Do persuasive cues such as urgency, authority framing, or social proof increase trust or reduce the influence of disclaimers?

\textbf{RQ4: AI-Specific Disclaimer Response.} Do AI-specific disclaimers reduce trust in AI-generated information, or do users continue to perceive AI responses as reliable despite warnings about inaccuracy or incompleteness?

Based on these questions, the study tests four hypotheses. \textbf{H1:} More visually prominent disclaimers, particularly end-bold and top-banner formats, will produce higher reported disclaimer influence than end-small or no-disclaimer conditions. \textbf{H2:} Trust and perceived correctness will vary by domain, with medicine/wellness expected to receive higher baseline trust than finance or AI-generated information. \textbf{H3:} Persuasion cues will increase trust and reduce reported disclaimer influence compared with neutral, persuasion-absent conditions. \textbf{H4:} AI-specific disclaimers will not fully reduce trust in AI-generated information, and some users may continue to report high trust despite explicit warnings.

\section{Data and Methodology}

This section describes the survey dataset, experimental design, stimulus construction, outcome measures, and analysis procedure used to examine how disclaimer placement and persuasive cues affect trust in finance, medicine/wellness, and AI-generated information.

\subsection{Dataset}

The study used survey data collected from 52 unique respondents. Each participant evaluated approximately ten advisory stimuli drawn from a larger set of 24 experimental conditions. After excluding open-ended final-question rows, the final dataset contained 378 usable stimulus-level responses. The individual stimulus-level response was used as the primary unit of analysis because each participant contributed ratings across multiple conditions.

The dataset included three domains: finance, medicine/wellness, and AI-generated information. Responses were also distributed across four disclaimer placement conditions and two persuasion cue conditions, as summarized in Table~\ref{tab:response_distribution}.

\begin{table}[t]
\centering
\caption{Stimulus-Level Response Distribution}
\label{tab:response_distribution}
\footnotesize
\begin{tabular}{llr}
\toprule
\textbf{Category} & \textbf{Level} & \textbf{Responses} \tabularnewline
\midrule
Domain & AI-generated information & 153 \tabularnewline
Domain & Medicine/Wellness & 113 \tabularnewline
Domain & Finance & 112 \tabularnewline
\midrule
Placement & No disclaimer & 99 \tabularnewline
Placement & End-small & 99 \tabularnewline
Placement & End-bold & 86 \tabularnewline
Placement & Top-banner & 94 \tabularnewline
\midrule
Persuasion & Absent & 189 \tabularnewline
Persuasion & Present & 189 \tabularnewline
\bottomrule
\end{tabular}
\end{table}

Basic demographic information was collected at the beginning of the survey using three dropdown fields: age group, gender, and highest level of education completed. Among the 52 respondents, the most common age groups were 18-24 and 25-34. The sample was predominantly male and highly educated, with most respondents reporting a Master's degree. These characteristics are considered when interpreting the findings because participants may have had higher digital literacy and greater familiarity with AI-generated content than the general population.

\subsection{Experimental Design}

The study used a mixed within-between experimental survey design organized around a 3~$\times$~4~$\times$~2 factorial structure. The three factors were domain, disclaimer placement, and persuasion cue condition. Crossing these factors produced 24 total experimental conditions, as shown in Table~\ref{tab:design}.

\begin{table}[t]
\centering
\caption{Experimental Factors and Levels}
\label{tab:design}
\footnotesize
\begin{tabular}{p{0.33\columnwidth}p{0.58\columnwidth}}
\toprule
\textbf{Factor} & \textbf{Levels} \tabularnewline
\midrule
Domain & Finance; Medicine/Wellness; AI-generated information \tabularnewline
Disclaimer Placement & No disclaimer; End-small; End-bold; Top-banner \tabularnewline
Persuasion Cue & Absent; Present \tabularnewline
\bottomrule
\end{tabular}
\end{table}

Participants did not evaluate only one condition. Instead, each participant viewed approximately ten stimuli drawn from across the full condition set. This repeated-measures structure allowed participants to contribute responses to multiple conditions while the survey platform balanced exposure across the overall stimulus set. Therefore, the design is best characterized as a mixed within-between design rather than a strictly between participants design.

\subsection{Stimuli and Disclaimer Conditions}

The study used 24 short advisory-style stimuli across finance, medicine/wellness, and AI-generated information. Each stimulus presented general informational guidance in a neutral, educational tone and avoided specific or actionable professional advice. Initial drafts were generated with the assistance of a large language model (GPT-4, early 2026 release) and then manually reviewed to balance length, reading level, and persuasive strength across domains.

During pilot refinement, stimuli judged to be too long were shortened to approximately 80--110 words, and the red-bordered top-banner disclaimer box was removed because it was visually distracting. The pilot confirmed that the stimuli were legible, the disclaimer placements were identifiable, and the survey could be completed in under ten minutes.

Disclaimer placement had four conditions: no disclaimer, end-small, end-bold, and top-banner. In the no-disclaimer condition, no advisory warning appeared. In the end-small and end-bold conditions, the disclaimer appeared after the paragraph in standard or bold text, respectively. In the top-banner condition, the disclaimer appeared before the paragraph as a visually prominent warning.

Disclaimer wording was standardized by domain. Finance stimuli used: ''This is general information, not financial advice.'' Medicine/wellness stimuli used: ''This is general information, not medical advice.'' AI stimuli used: ''This is general information about AI systems and may be inaccurate or incomplete.'' The complete set of 24 stimuli is available as supplementary material.

Persuasion-present stimuli included social proof, authority framing, or urgency, using phrases such as ``millions of people,'' ``experts recommend,'' and ``don't wait.'' Persuasion-absent stimuli remained neutral and informational without these cues.

\subsection{Survey Procedure and Measures}

Participants first completed a brief demographic screen asking for age group, gender, and highest level of education completed. They then evaluated approximately ten stimuli, each presented on a separate screen under a ``Context'' heading and followed by three rating questions.

Three quantitative dependent variables were recorded for each stimulus. \textit{Trust} (Q1) asked: `Rate your level of trust in the information presented in this paragraph.'' Responses ranged from 1 = No trust to 5 = Complete trust. \textit{Likelihood} (Q2) asked: `Rate the likelihood that the information in this paragraph is correct.'' Responses ranged from 1 = Very unlikely to be correct to 5 = Very likely to be correct. \textit{Disclaimer Influence} (Q3) asked: ``Rate how much the disclaimer influenced your opinion.'' Responses ranged from 1 = No influence at all to 5 = Very strong influence.

After the stimulus-evaluation task, participants answered three open-ended reflection questions about AI trust, responsibility for harm, and reliance on AI-generated information without ethical or safety guidance. These responses were analyzed qualitatively to support interpretation of the quantitative findings.

\subsection{Analysis Plan}

The quantitative analysis focused on trust, perceived likelihood that the information was correct, and disclaimer influence. To rigorously account for the nested data structure—since each participant contributed multiple responses—we employed linear mixed-effects models (LMM) with participant-level random intercepts. This approach robustly controls for individual baseline differences in trust and disclaimer engagement. One-way analysis of variance (ANOVA) was also utilized for descriptive baseline comparisons across domains and placement conditions.

\subsection{Ethics and Safety}

The study was designed to minimize participant risk. All advisory paragraphs were written in general, non-specific language and did not provide individualized medical, financial, or AI-usage advice. No participant was instructed to act on any information presented in the stimuli.

No personally identifiable information was collected. Participants were not asked for names, email addresses, financial account details, health records, or specific AI usage histories. Participation was voluntary, and informed consent was obtained before study materials were displayed. At the end of the survey, participants received a debrief explaining that the advisory paragraphs were simulated experimental stimuli and were not actual sources of financial, medical, or AI guidance.

\section{Results}

This section reports the main quantitative findings across the three outcome measures: trust, perceived likelihood that the information was correct, and disclaimer influence.

\subsection{Overall Response Patterns}

Across 378 stimulus-level responses, participants reported generally high trust in the advisory paragraphs. Mean trust was $M = 3.94$ ($SD = 0.86$) on a 1-5 scale, with the modal response being ``4 - High trust'' ($n = 193$, 51.1\%). Perceived likelihood that the information was correct was slightly higher, with $M = 4.04$ ($SD = 0.87$). Together, these results suggest that participants generally viewed the advisory content as credible regardless of disclaimer condition.

Disclaimer influence showed greater variation than trust or likelihood. The mean disclaimer influence rating was $M = 3.25$ ($SD = 1.26$), indicating moderate influence overall. Responses were distributed across both low-influence and high-influence categories, suggesting that participants differed substantially in how much they reported being affected by disclaimers.

\begin{figure}[t]
\centering
\includegraphics[width=\columnwidth]{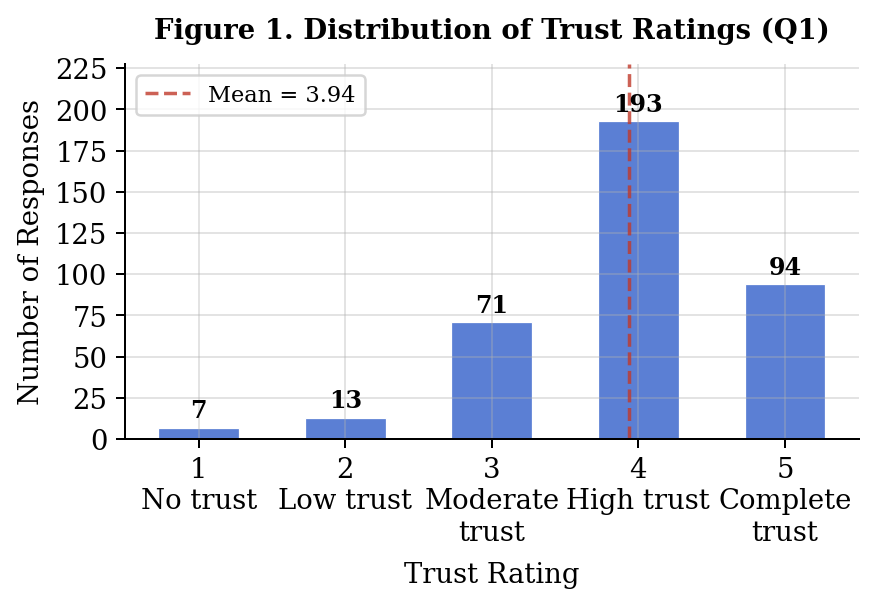}
\caption{Distribution of trust ratings (Q1) across all 378 stimulus-level responses. Mean trust was 3.94 on a 1-5 scale.}
\label{fig:trust_dist}
\end{figure}

\subsection{Disclaimer Placement and Reported Influence}

To address RQ1, we examined whether disclaimer placement affected reported disclaimer influence. Mean disclaimer influence was lowest in the no-disclaimer condition ($M = 3.12$, $SD = 1.39$) and highest in the end-bold condition ($M = 3.40$, $SD = 1.14$). The end-small condition produced a mean of $M = 3.23$ ($SD = 1.19$), while the top-banner condition produced a mean of $M = 3.26$ ($SD = 1.28$). Although the means followed a generally expected pattern in which visually emphasized disclaimers produced higher reported influence, the differences were small.

A linear mixed-effects model predicting disclaimer influence with a random intercept for participant found no statistically significant effect of placement condition. Thus, H1 received only partial descriptive support. Bold end-positioned disclaimers produced the highest mean influence, but disclaimer placement did not significantly change reported influence in this sample once individual participant variance was accounted for.

\begin{figure}[t]
\centering
\includegraphics[width=\columnwidth]{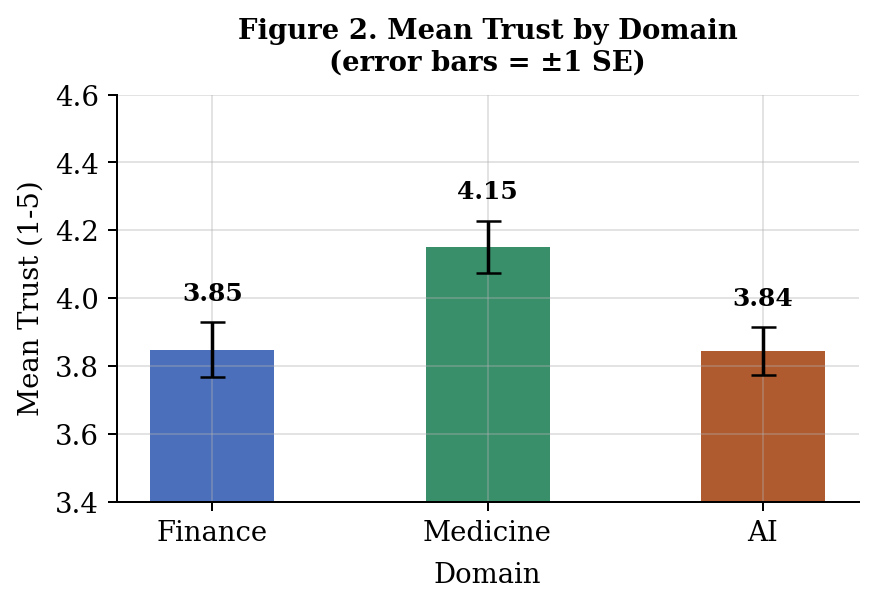}
\caption{Mean trust (Q1) by domain. Error bars represent $\pm$1 SE. Medicine/wellness content received the highest trust ratings.}
\label{fig:place}
\end{figure}

\subsection{Domain Differences in Trust}

To address RQ2, we compared trust ratings across finance, medicine/wellness, and AI-generated information. Trust differed meaningfully by domain. Medical content received the highest trust ratings ($M = 4.15$, $SD = 0.83$), followed by finance ($M = 3.85$, $SD = 0.85$) and AI-generated information ($M = 3.84$, $SD = 0.87$). A linear mixed-effects model confirmed a statistically significant domain effect on trust, accounting for participant-level variance ($p < .01$).

This result indicates that domain context shaped how participants evaluated advisory content. Medical/wellness paragraphs were trusted more than finance and AI paragraphs, even though all domains used the same general experimental structure. Perceived likelihood followed the same pattern, with medicine receiving the highest mean likelihood rating ($M = 4.21$), followed by AI ($M = 3.99$) and finance ($M = 3.95$).

\begin{figure}[t]
\centering
\includegraphics[width=\columnwidth]{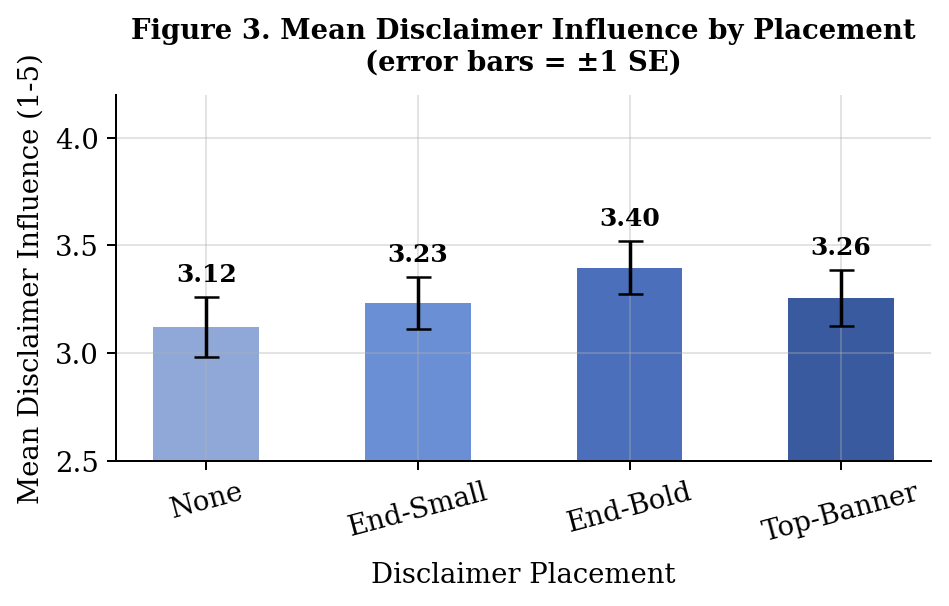}
\caption{Mean disclaimer influence (Q3) by placement condition. Error bars represent $\pm$1 SE. End-bold placement produced the highest mean influence, but the placement effect was not statistically significant.}
\label{fig:domain}
\end{figure}

\subsection{Persuasion Cues and Disclaimer Engagement}

To address RQ3, we compared persuasion-present and persuasion-absent conditions. Mean disclaimer influence was nearly identical across the two groups: persuasion-absent conditions produced $M = 3.23$ ($SD = 1.31$), while persuasion-present conditions produced $M = 3.26$ ($SD = 1.20$). This difference was negligible, suggesting that persuasion cues did not meaningfully change reported disclaimer influence.

Trust also did not increase in the presence of persuasion cues. Contrary to H3, mean trust was slightly higher in the persuasion-absent condition ($M = 3.97$) than in the persuasion-present condition ($M = 3.90$). This pattern suggests that cues such as urgency, authority, and social proof did not make the paragraphs more trusted in this sample. One possible explanation is that the highly educated and technically literate participant group may have recognized these cues as persuasive or promotional language, reducing rather than increasing trust.

\begin{figure}[t]
\centering
\includegraphics[width=\columnwidth]{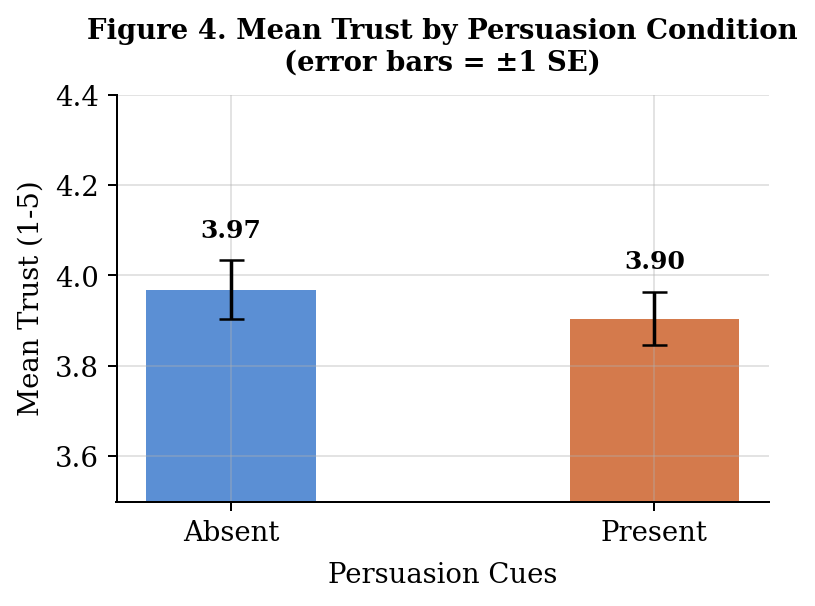}
\caption{Mean trust (Q1) by persuasion cue condition. Error bars represent $\pm$1 SE. Persuasion-present conditions did not increase trust.}
\label{fig:persuasion}
\end{figure}

\subsection{AI-Specific Disclaimer Response}

To address RQ4, we examined whether AI-specific disclaimers reduced overconfidence in AI-generated information. The AI domain produced the lowest mean trust score ($M = 3.84$) and the lowest mean likelihood rating ($M = 3.99$) of the three domains. However, these values remained above the midpoint of the scale, suggesting that participants still viewed AI-generated content as generally credible even when disclaimers warned that the information could be inaccurate or incomplete.

Disclaimer influence in the AI domain ($M = 3.24$) was nearly identical to finance ($M = 3.27$) and medicine/wellness ($M = 3.23$). This indicates that AI-specific disclaimers did not produce stronger caution than disclaimers in the comparison domains. Qualitative responses further suggested a transparency paradox: some participants interpreted AI disclaimers as signs of honesty or system self-awareness rather than as warnings. For example, one respondent stated that the disclaimer made the AI seem aware of its limitations, which increased confidence in the information.

The AI domain also showed evidence consistent with banner blindness. Top-banner placement produced the lowest mean disclaimer influence in the AI condition ($M = 3.11$), lower even than the no-disclaimer condition ($M = 3.12$). This suggests that users familiar with AI interfaces may process repeated top-positioned warnings as routine background text rather than as meaningful risk signals. Overall, the AI results provide mixed support for H4: participants did not show uniquely high trust in AI content, but AI disclaimers also did not substantially reduce trust or perceived correctness.

\section{Discussion}
The results suggest that disclaimers alone had limited ability to reduce trust in advisory information. Across conditions, participants generally rated the stimulus paragraphs as credible, while reported disclaimer influence remained moderate. This indicates that disclaimers may not function as strong safeguards when the surrounding content appears clear, useful, or authoritative. Their effect appears to depend on domain context, presentation format, and user interpretation.

\subsection{Domain and Placement Effects}
The significant domain effect shows that users do not evaluate disclaimers in isolation. Medicine/wellness content received higher trust ratings than finance or AI-generated information, suggesting that users bring prior expectations about expertise, risk, and authority into their evaluation of advisory content. A disclaimer attached to medical information may therefore be interpreted differently from one attached to financial or AI-generated information.
Disclaimer placement had a weaker effect. Although end-bold disclaimers produced the highest mean influence, the differences across placement conditions were not statistically significant in the mixed-effects model. This suggests that visual emphasis may help users notice a warning, but placement alone is unlikely to meaningfully change trust when the main message already appears credible.

\subsection{Persuasion Cues and User Skepticism}
Persuasion cues did not increase trust as predicted. One possible explanation is that the sample, which was highly educated and likely familiar with digital and AI-generated content, recognized cues such as urgency, authority framing, and social proof as persuasive techniques. Instead of increasing credibility, these cues may have made some paragraphs appear promotional or less neutral.
This finding has implications for AI-generated communication. Confidence-building language may not always increase trust, especially among critical users. Responsible AI design should therefore prioritize clarity, evidence, and uncertainty communication over persuasive style.

\subsection{AI Disclaimers and Behavioral Psychology}
The AI-specific findings suggest a transparency paradox. AI disclaimers warned that information could be inaccurate or incomplete, yet they did not substantially reduce trust. Some participants interpreted such warnings as signs of honesty, self-awareness, or transparency rather than caution. This aligns with behavioral psychology concepts such as the \textit{halo effect}, where a system's explicit declaration of its own limitations makes it appear more ethical and reliable overall. Paradoxically, this can reinforce \textit{automation bias}—the well-documented tendency for users to favor machine-generated outputs and heuristics even when explicitly warned of potential flaws. This confirms prior work showing that transparency features in AI systems do not always reduce reliance and may sometimes increase perceived trustworthiness~\cite{benda2021appropriatetrust, scharowski2023explanations, henestrosa2025aidisclaimers}.
The AI results also suggest possible banner blindness. Top-banner disclaimers produced the lowest mean influence within the AI condition, indicating that repeated, standardized warnings may be treated as routine interface elements rather than meaningful risk signals. Over time, repeated exposure may weaken attention and reduce the protective effect of the warning~\cite{cranor2008humanwarnings}.

\subsection{Design, Fairness, and Accountability Implications}
These findings suggest that static disclaimers are not enough for responsible AI design. Because our highly educated sample frequently ignored disclaimers or interpreted them as transparency, there are critical implications for algorithmic bias and fairness. If highly digitally literate users fail to calibrate their trust appropriately, these static disclaimers are likely even less protective for vulnerable populations with lower digital literacy. This disparity highlights a significant fairness issue: standardized disclaimers may disproportionately fail to protect those most at risk of being harmed by inaccurate AI-generated guidance.

Rather than relying only on generic disclaimers, AI developers and platforms must design warnings that are visible, context-specific, and proportionate to the risk of the output. High-stakes contexts such as health, finance, and professional decision-making may require dynamic warnings tied to specific claims, stronger verification prompts, and clearer signals when information is unsupported. Disclaimer design must be treated as both a communication and accountability issue.

\section{Limitations and Future Work}

Several limitations should be considered. First, the sample was small and non-representative. The 52 respondents were predominantly young, highly educated, and technically literate, which may limit generalizability while simultaneously highlighting the inefficacy of disclaimers even among sophisticated populations. A larger and more diverse sample is needed to test whether the same patterns hold across different literacy levels and AI familiarity.

Second, the study relied on self-reported ratings. Trust, perceived correctness, and disclaimer influence were measured using 1-5 scales, but the study did not directly measure whether participants visually noticed or read the disclaimer. Future studies should include behavioral measures such as eye-tracking, reading time, verification behavior, or decision-based tasks.

Third, while we utilized mixed-effects models to account for the repeated-measures structure of the data, the experimental setting remains simulated. Future work should also measure prior familiarity with AI systems, finance, medicine, and online risk disclosures in ecologically valid environments. Future experiments should compare static disclaimers with dynamic, claim-specific warnings and interactive verification prompts, as well as examine whether repeated exposure increases disclaimer habituation over time.

\section{Conclusion}
This study examined how disclaimer placement and persuasive cues shape trust, perceived correctness, and reported disclaimer influence across finance, medicine/wellness, and AI-generated information. Across 378 stimulus-level responses from 52 participants, advisory content was generally rated as credible, while disclaimers showed only moderate influence. Domain context mattered, with medicine/wellness content receiving the highest trust ratings. Although end-bold disclaimers produced the highest mean influence, placement effects were not statistically significant, and persuasion cues did not increase trust as expected. The AI findings suggest that generic disclaimers may be insufficient because some users interpret them as signs of transparency rather than caution, while repeated top-positioned warnings may contribute to banner blindness. Overall, responsible AI design should move beyond static disclaimers toward context-specific risk communication that helps users evaluate uncertainty more effectively.

\section{Acknowledgment}
The author thanks Lavanya Prahallad for providing research guidance and mentorship during am internship. This work culminated in the publication of this research paper.

% ── References ────────────────────────────────────────────────────────────────
\bibliographystyle{IEEEtran}
\bibliography{references}
\end{document}